%% file: main.tex
\documentclass[journal]{IEEEtran}
\usepackage{cite}
\usepackage{graphicx} % Required for 
\usepackage{amsmath,amsfonts,amssymb}
\usepackage{url}

\usepackage{tikz}
\usepackage[caption=false,font=normalsize]{subfig}
\input{math_variables}
\title{Adaptive Experiment Design for Nonlinear System Identification with Operational Constraints}
\author{Jingwei Hu, Dave Zachariah, Torbj\"{o}rn Wigren, \mbox{Petre Stoica}
\thanks{All authors are with Dept. of Information Technology, Uppsala University, Sweden}}

\begin{document}
\maketitle

\begin{abstract}       
We consider the joint problem of online experiment design and parameter estimation for identifying nonlinear system models, while adhering to system constraints. We utilize a receding horizon approach and propose a new adaptive input design criterion, which is tailored to continuously updated parameter estimates, along with a new sequential estimator. We demonstrate the ability of the method to design informative experiments online, while steering the system within operational constraints. 
\end{abstract}

\section{Introduction}

We consider discrete-time nonlinear dynamical systems of the form
\begin{equation}
\stats_{\ts+1} = f_{\modelpara}(\stats_{\ts},\inp_{\ts}),
\label{eq:nonlinearsystem}
\end{equation}
where both states $\stats_{\ts}$ and parameters $\modelpara$ are unknown. We aim to address the joint problem of (i) \emph{estimating} $\modelpara$ from the inputs $\inp_{\ts}$ and measurements of the system outputs $\oup_\ts$, and (ii) \emph{designing} $\inp_{\ts}$ so as to make the data informative about $\modelpara$, while steering the system states within operational constraints. 

 There is a rich set of methods to tackle the estimation and design problems separately when the system is known to be \emph{linear}. A classic design choice for $\inp_{\ts}$ is a pseudorandom binary sequence (\prbstext{}) \cite{godfrey1991introduction} due to its excitation properties and ease of implementation. More efficient input designs are possible by leveraging a known structure of the system using the statistically motivated Fisher information matrix or asymptotic error covariance matrices, in the design criterion \cite{kiefer1959optimum,federov1972optimalexp,mehra1974optimal,goodwin1977dynamic,stoica1982useful, bombois2011optimal,jansson2005input,eckhard2013input,annergren2017application}.
 
 The high cost of conducting  experiments motivates the development of efficient design schemes that continuously \emph{adapt} $\inp_{\ts}$ to the information available up to time $\ts$ \cite{arimoto1971optimum,keviczky1972input,goodwin1977dynamic,stojanovic2014adaptive,mania2022active}.  Informative experiments may, however, have adverse effects on the system in operation. (See \cite{feldbaum1965optimal,flila2008optimal,qian2017optimal} for a control perspective.) An open-loop input design method for \emph{nonlinear} systems was introduced in \cite{goodwin1971optimal}, that penalizes designs in which the estimated states violate the constraints. This is achieved by linearizing the state-space model and designing $\inp_{\ts}$ over a fixed horizon. More recently, a closed-loop (receding horizon) method was developed in \cite{babar2021optimal} that also penalizes constraint violations. This penalized design approach requires a delicate balancing of competing objectives, where the  information criterion is based on direct combination of dimensions of $\modelpara$ that typically have incommensurable units. 
 
 %TODO In parallel, \cite{mania2022active} studies active data collection for learning nonlinear dynamical systems, adaptively exciting poorly explored feature directions under input-magnitude constraints and deriving finite-time identification guarantees.

The contribution of this paper is: 
\begin{itemize}
    \item a statistically motivated experiment design method,  adapted to the available information at each time step, that can steer the system within operational constraints; and
   \item an online parameter estimation method that is efficient and insensitive to initial estimates.
\end{itemize}
The proposed joint design and estimation method enables safer and more efficient experiments for nonlinear systems.

\section{Problem Formulation}

The system output is assumed to follow the measurement model
\begin{equation}
    \oup_{\ts} = \regressormatrix\stats_{\ts} + e_{\ts}
    \label{eq:linearobservation}
\end{equation} 
where $\oup_{\ts}$ is a $d_{\oup} \times 1$ vector and $e_{\ts} \sim \mathcal{N}(0, \noisecov)$ is a white noise process with a diagonal covariance matrix $\noisecov$ with diagonal elements given by the vector $\noisev$. Sensor specifications typically provide relatively accurate estimates of the variances $\noisev$. The dimensions of $\inp_{\ts}$, $\modelpara$ and $\stats_{\ts}$ are $d_{\inp} \times 1$,  $d_{\modelpara} \times 1$ and $d_{\stats} \times 1$, respectively.

A sequence of input-output signals $\{(\inp_{0},\oup_{1}),\dots, (\inp_{\ts-1},\oup_{\ts})\}$ provides information about $\modelpara$ in \eqref{eq:nonlinearsystem} via \eqref{eq:linearobservation}. At any $\ts$, our aim is to design subsequent inputs $\inp_{\ts+i}$ so that the resulting observations provide the maximal \emph{additional} information about $\modelpara$, while steering the system within user-defined operational constraints:
\begin{equation}
\stats_{\min}\leq \stats_{\ts+i} \leq \stats_{\max},
\label{eq:stateconstraints}
\end{equation}
which can be time varying. This is readily extended to constraints on the outputs $\regressormatrix\stats_{\ts}$ via \eqref{eq:linearobservation}. A fundamental challenge here is that $\stats_{\ts+i}$ is given by the forward recursion of \eqref{eq:nonlinearsystem}:
\begin{equation}
\begin{split}
\stats_{\ts+i} &= f_\modelpara(f_\modelpara(\dots f_\modelpara(\stats_{\ts},\inp_{\ts})\dots\inp_{\ts+i-2}),\inp_{\ts+i-1})\\
&\equiv f^{i}_{\modelpara}(\stats_{\ts},\inp_{\ts:\ts+i-1}), \; \forall i>0
\end{split}
\label{eq:stateforwardrecursion}
\end{equation}
where $\modelpara$ and $\stats_{\ts+1}$ have to be estimated from data. Thus \eqref{eq:stateforwardrecursion} renders the constraints \eqref{eq:stateconstraints} nonlinear in the designed inputs, which must also satisfy
\begin{equation}
    \inp_{\min}\leq \inp_{\ts+i} \leq \inp_{\max}.
\label{eq:inputconstraints}
\end{equation}

Let $\cinp$ denote the set of input sequences that satisfy \eqref{eq:inputconstraints} and let $\cstats$ be the set of resulting states that satisfy \eqref{eq:stateconstraints}. It is assumed that a feasible sequence exists in $\cinp$. To tackle the computational challenges ensuing from \eqref{eq:stateforwardrecursion}, we will leverage advances in automatic differentiation methods. 

\emph{Notation:} $\| z \|_{W} = \sqrt{ z^\top W z }$ is a weighted norm using a positive definite matrix $W$. The element-wise product and division operator are denoted $\odot$ and $\oslash$, while $(z)^{+}$ sets the negative elements of a vector $z$ to 0, i.e., $[(z)^+]_i = \max(0,z_i)$.

\section{Method}

Suppose we are to design a $k$-step input sequence 
$\kinp_{\ts} = [ \inp_{\ts}, \inp_{\ts+1}, \dots, \inp_{\ts+\hs-1}] $. The horizon length $k$ should be set sufficiently long to steer the system within operational constraints. The designed sequence yields corresponding outputs $\{\oup_{t+1}, \dots, \oup_{\ts+\hs} \}$.   We begin by formulating a suitable adaptive design criterion for $\kinp_{\ts}$ that quantifies the additional information gained about $\modelpara$ relative to what was available at $\ts$.

\subsection{Adaptive design criterion}

Using \eqref{eq:linearobservation} and \eqref{eq:stateforwardrecursion}, the likelihood of the $\hs$ new outputs depends on $\modelpara$, $\noisev$ and the unknown state $\stats_{\ts}$. Implementing $\kinp_{\ts}$ yields the following Fisher information matrix (\fimtext{}) of the unknown variables:
\begin{equation}
\FIM_{\modelpara,\stats_{\ts},\noisev}(\kinp_{\ts}) = \expectation\left[\begin{bmatrix}
   \frac{\partial \loglikelihood}{\partial \modelpara}\\
    \frac{\partial \loglikelihood}{\partial \stats_{\ts}}\\
    \frac{\partial \loglikelihood}{\partial \noisev}\\
\end{bmatrix} 
\begin{bmatrix}\frac{\partial \loglikelihood}{\partial \modelpara}\\
    \frac{\partial \loglikelihood}{\partial \stats_{\ts}}\\
    \frac{\partial \loglikelihood}{\partial \noisev}\\
\end{bmatrix}^\top \right],
\label{eq:jointFIM}
\end{equation}
where $\loglikelihood$ denotes the (Gaussian) log-likelihood function of the data. It follows from the Cram\'{e}r-Rao inequality \cite{soderstrom1989system,ljung1999system} that
\begin{equation}
\begin{split}
\errorcov_{\ts+k} = \begin{bmatrix}
I_{d_{\modelpara}} & 0
\end{bmatrix} \FIM^{-1}_{\modelpara,\stats_{\ts},\noisev} 
\end{split} \begin{bmatrix}
I_{d_{\modelpara}} & 0
\end{bmatrix}^\top
\label{eq:crbinequality}
\end{equation}
is the minimum achievable error covariance matrix for any unbiased estimator $\modelparaest_{\ts+k}$ that incorporates the $k$ new samples.

Let $\modelparaest_{\ts}$ denote a prior estimate at time $\ts$ with error covariance matrix $\errorcov_{\ts} \succeq \errorcov_{\ts+k}$. Then the following bounded criterion
\begin{equation}
\boxed{\cost(\kinp_{\ts}; \modelpara, \stats_{\ts}, \noisev) = \text{tr}\left\{ \errorcov_{\ts+k} \errorcov^{-1}_\ts \right\} \leq d_{\modelpara} ,}
\label{eq:weightedtrace}
\end{equation}
quantifies the \emph{relative} error improvement gained by the design $\kinp_{\ts}$. The upper bound $d_{\modelpara}$ is attained in the extreme case of \emph{uninformative} experiments. The criterion \eqref{eq:weightedtrace}  bounds the mean squared-error of  $\modelparaest_{\ts+k}$, standardized to the errors at $\ts$:
\begin{equation}
\cost \leq \expectation\left[\| \modelpara - \modelparaest_{\ts+k}\|^2_{\errorcov^{-1}_{\ts}}  \right],
\label{eq:crb}
\end{equation}
via the Cram\'{e}r-Rao inequality. Thus $\cost$ is an L-criterion \cite[ch.~2.9]{federov1972optimalexp}  \emph{adapted} to the information about $\modelpara$ available at $\ts$ that renders the errors in the components of $\modelpara$ commensurable.

A receding horizon formulation for adaptive experiment design is then
\begin{equation}
\begin{split}
\kinp^*_{\ts} &= \argmin_{\kinp_{\ts} \in \cinp} \cost(\kinp_{\ts}; \modelpara, \stats_{\ts}, \noisev) \\
&\; \text{subject to } \stats_{\ts+i} \in \cstats, \forall i=1,\dots,k,
\end{split}
\label{eq:optimize-oc}
\end{equation}
which is evaluated at estimates $\modelparaest$, $\statsest_{\ts}$ and $\noisevest$ given in Section~\ref{sec:estimator}. However, both the constraints in \eqref{eq:optimize-oc} and the log-likelihood for all data render the problem rather intractable. Next, we address both issues.

\subsection{Relaxation}

The likelihood function for the past data $\oup_{1:\ts}$ does not depend on $\kinp_{\ts}$ and is parameterized using an initial state $\stats_0$. To obtain a tractable \fimtext{} that uses all data, we consider summarizing the past data $\oup_{1:\ts}$ using maximum likelihood estimates (\mletext{}) $\modelparaest, \statsest_{\ts}$ and $ \noisevest$. The distribution of the estimates is denoted $q(\modelparaest,\statsest_{\ts},\noisevest|\modelpara,\stats_{\ts},\noisev)$. Under standard regularity conditions \cite[ch.~9]{ljung1999system}\cite[app.~B]{stoica2005spectral}, $q$ is asymptotically normal and the estimator efficient so that with a Gaussian data model \eqref{eq:linearobservation}, its asymptotic covariance matrix is block-diagonal:
\begin{equation}
\mathbb{V}\left[
\begin{bmatrix}
\modelparaest &
\statsest_{\ts} &
\noisevest 
\end{bmatrix}^\top \right] = 
\begin{bmatrix}
\errorcovjoint & 0 \\
0 & \errorcovv    
\end{bmatrix},
\label{eq:priorestimatecov}
\end{equation}
where the lower block $\errorcovv$ belongs to $\noisevest$ which is thus uncorrelated with $\modelparaest$ and $\statsest_{\ts}$. We will therefore take $q$ to be a Gaussian distribution as summary of the past data. With a given covariance matrix \eqref{eq:priorestimatecov}, this is the most conservative distribution in the sense that it yields the smallest \fimtext{} \cite{stoica2011gaussian,park2013gaussian}. (When the \mletext{} is asymptotically sufficient \cite{kaufman1966asymptotic}, $q$ also corresponds to a factorization of the past likelihood of $y_{1:t}$.)

The likelihood function then becomes
\begin{equation}   
\begin{split}
\loglikelihoodapprox &= \ln q(\modelparaest,\statsest_{\ts},\noisevest|\modelpara,\stats_{\ts},\noisev) + \sum^k_{i=1} p(\oup_{\ts+i}| \inp_{t:t+i-1}, \modelpara, \stats_{\ts}, \noisev) \\
&=-\frac{1}{2}\left\|\begin{bmatrix}
    \modelpara\\
    \stats_{\ts} \\
\end{bmatrix}-\begin{bmatrix}
    \modelparaest\\
    \statsest_{\ts} \\
\end{bmatrix}\right\|_{\errorcovjoint^{-1}}^2 -\frac{1}{2}\left\|    \noisev-
    \noisevest\right\|_{\errorcovv ^{-1}}^2 \\
&\quad - \frac{1}{2}\sum_{i=1}^{\hs}\|\prederr_{\ts+i}\|_{\noisecov^{-1}}^2 - \frac{\hs}{2}\ln{|\noisecov|}+ \text{const.}, \label{eq:mle_design}
\end{split}
\end{equation}
where we used the prediction error $\prederr_{\ts+i} \equiv \oup_{\ts+i} - \regressormatrix f^{i}_{\modelpara}(\stats_{\ts},\inp_{\ts:\ts+i-1})$ for compactness. The corresponding conservative \fimtext{} \eqref{eq:jointFIM} becomes \cite{soderstrom1989system}:
\begin{equation}
\begin{split}
    \FIMapprox_{\stats_{\ts},\modelpara,\noisev} &= \begin{bmatrix}\FIMapprox_{\modelpara}&\FIMapprox_{\modelpara\stats_{\ts}}&0\\
\FIMapprox_{\modelpara\stats_{\ts}}^{\top}&\FIMapprox_{\stats_{\ts}}&0\\
0&0&\FIMapprox_{\noisev}\end{bmatrix} \\
&= \begin{bmatrix}\errorcovjoint^{-1}+\sum_{i=1}^{\hs}\prederrgrad^{\top}_{\ts+i} \noisecov^{-1} \prederrgrad_{\ts+i}&{0}\\
    {0}&\errorcovv^{-1} + \frac{\hs}{2} \noisecov^{-2}\end{bmatrix},
\end{split}\label{eq:fim_block}
\end{equation}
where 
$$\prederrgrad_{\ts+i} = \regressormatrix\begin{bmatrix}
    
\frac{\partial f^{i}_{\modelpara}(\stats_{\ts},\inp_{\ts:\ts+i-1})}{\partial \modelpara}&
\frac{\partial f^{i}_{\modelpara}(\stats_{\ts},\inp_{\ts:\ts+i-1})}{\partial \stats_{\ts}}\end{bmatrix}$$
is the Jacobian of the prediction error $\prederr_{\ts+i}$. Applying \eqref{eq:fim_block} to \eqref{eq:crbinequality}, we finally obtain 
\begin{equation}
    \errorcovapprox_{\ts+\hs} \equiv ( \FIMapprox_{\modelpara} - \FIMapprox_{\modelpara\stats_{\ts}}\FIMapprox_{\stats_{\ts}}^{-1}\FIMapprox_{\modelpara\stats_{\ts}}^{\top} )^{-1} \succeq \errorcov_{\ts+\hs} .\label{eq:fim}
\end{equation}
Inserting the matrix in \eqref{eq:fim} into the criterion \eqref{eq:weightedtrace} yields a tractable, conservative minimum error bound $\widetilde{\cost} \geq \cost$. (The bound could be loose for small $t$, but can become tight when asymptotic sufficiency of the \mletext{} holds). The dominant cost of evaluating $\widetilde{\cost}$ is therefore the inversion of $\FIMapprox_{\stats_{\ts}}$. 

We now turn to relaxing the state constraints $\cstats$ in \eqref{eq:optimize-oc}. We aim for designs that continually steer the state to satisfy the operational constraints. This is achieved by introducing a penalty term to the design problem \eqref{eq:optimize-oc}. Whereas $\cost$ measures a standarized squared deviation  of $\modelparaest_{\ts+k}$ from the unknown $\modelpara$, we let penalty $\mathcal{J}_{\cstats}$ measure a standardized squared exceedance of $\stats_{\ts + i}$ from the state boundaries. Specifically,
\begin{equation}
\begin{split}
    &\mathcal{J}_{\cstats}(\kinp_{\ts}; \modelpara, \stats_{\ts}) = \frac{1 }{k} \times \\
     &\sum_{i=1}^{\hs}  \|(S\stats_{\min}-S\stats_{\ts+i})^{+} + (S\stats_{\ts+i}- S\stats_{\text{max}})^{+})\|^2,
\end{split}\label{costoperationalconstraints}
\end{equation}
where $S$ is a diagonal matrix that standardizes violations from the constraints. That is, $[S]_{i,i}=1/(\stats^{(i)}_{\max} - \stats^{(i)}_{\min})$ or 0, depending on whether $\stats_i$ is constrained. 
The relaxed form of \eqref{eq:optimize-oc} then becomes
\begin{equation}
\kinp^*_{\ts} = \argmin_{\kinp_{\ts} \in \cinp} \widetilde{\cost}(\kinp_{\ts}; \modelpara, \stats_{\ts}, \noisev) + \gamma \mathcal{J}_{\cstats}(\kinp_{\ts}; \modelpara, \stats_{\ts}),
\label{eq:optimize-relax}
\end{equation} 
where $\gamma \geq 0$ is a weight that controls the adherence to the state constraints. Note that when the state sequence $\stats_{\ts+i}$ is in $\cstats$, the cost $\mathcal{J}_{\cstats}$ vanishes. By setting, for instance, $\gamma > d_{\modelpara}/(0.01)^2$ the cost of exceeding the state boundaries by 1\%, is greater than the cost of using uninformative inputs ($\widetilde{\cost} = d_{\modelpara}$). This incentivizes designs that steer the system within the state constraint set $\cstats$. Note that $\widetilde{\cost}$ also provides a diagnostic for how informative the experiment is.

\subsection{Online estimator}
\label{sec:estimator}
Joint estimation of $(\modelpara,\stats_{\ts}, \noisev)$ for the design \eqref{eq:optimize-relax} involves $d_{\modelpara} + d_{\stats}+d_{\oup}$ parameters and is a challenging task for standard recursive estimators. To reduce the sensitivity to poor initial estimates, we update $\modelparaest$ periodically using blocks of $\blocksize \geq d_{\modelpara} + d_{\stats}+d_{\oup}$ samples. At the same time, we maintain a running estimation of $\stats_{\ts}$ using the model in \eqref{eq:nonlinearsystem}. 

Suppose $\ts$ is a multiple of $\blocksize$, so that $\tp=\ts-\blocksize$ indicates the end of the previous block of samples. Let $(\modelparaest_{\tp},\statsest_{\tp|\tp}, \noisevest_{\tp})$ denote the prior estimates at $\tp$. Using the likelihood function along with the $\blocksize$ subsequent samples, new estimates $\modelparaest_{\ts}$, $\noisevest_{\ts}$, and a \emph{smoothed} estimate $\statsest_{\tp|\ts}$ is obtained by solving:
\begin{equation}
\begin{split}
&(\modelparaest_{\ts},\statsest_{\tp|\ts}, \noisevest_{\ts}) 
= \argmin_{\modelpara,\stats_{\tp},\noisev}\; \sum_{i=1}^{\blocksize}\|\prederr_{\tp+i}\|_{\noisecov^{-1}}^2 + \blocksize\ln{|\noisecov|} \\ 
    &+\left\|\begin{bmatrix}
    \modelpara\\
    \stats_{\tp}
\end{bmatrix}-\begin{bmatrix}
    \modelparaest_{\tp}\\
    \statsest_{\tp|\tp}
\end{bmatrix}\right\|_{\errorcovjoint_{\tp|\tp}^{-1}}^2
    + \|\noisev - \noisevest_{\tp}\|^2_{\errorcovv^{-1}_{\tp}},
\end{split}
\label{eq:blockwiseestimator}
\end{equation}
which is a maximum likelihood estimator. The covariance matrices are therefore approximated using the corresponding inverse \fimtext{}:
\begin{equation}
    \errorcovjoint_{\tp|\ts} = \begin{bmatrix}\FIMapprox_{\modelpara}&\FIMapprox_{\modelpara\stats_{\tp}}\\
\FIMapprox_{\modelpara\stats_{\tp}}^{\top}&\FIMapprox_{\stats_{\tp}}
\end{bmatrix}^{-1}
\quad \text{and} \quad \errorcovv_{\ts} = \FIMapprox_{\noisev}^{-1} \label{eq:estcov}
\end{equation}
 based on the form in \eqref{eq:fim_block} with plug-in estimates. The state estimate in \eqref{eq:blockwiseestimator} and the joint covariance \eqref{eq:estcov} are then sequentially updated to $\ts$:
\begin{equation*}
(\statsest_{\tp|\ts},\errorcovjoint_{\tp|\ts}) \rightarrow (\statsest_{\tp+1|\ts},\errorcovjoint_{\tp+1|\ts}) \rightarrow \dots \rightarrow (\statsest_{\ts|\ts},\errorcovjoint_{\ts|\ts})
\end{equation*}
using the prediction step of the unscented Kalman filter  \cite{julier2004unscented,menegaz2015systematization,candy2016bayesian}. The procedure is described as follows:

Let $\augparaest_{\ts} = \begin{bmatrix}
    \modelparaest \\
    \statsest_{\ts} 
\end{bmatrix}$ denote an initial estimate with an error covariance matrix  $\errorcovjoint_{\ts}$. The unscented transform is a deterministic sampling method that produces a one-step prediction, and its approximate error covariance, using $S=2d_z + 1$ samples with $d_z = d_{\modelpara} + d_{\stats}$, is given by:
\begin{equation}
\begin{split}
(\augparaest_{\ts+1},\errorcovjoint_{\ts+1}) &=T(\augparaest_{\ts},\errorcovjoint_{\ts};\inp_\ts).
\end{split}
\label{eq:sigmapoint_transformation}
\end{equation}
The augmented state estimate is computed as
\begin{equation*}
    \augparaest_{\ts+1} = \frac{1}{S}\sum_{s=0}^{S-1} w^{(s)}\augpara^{(s)}_{\ts+1}, \text{ where } \augpara^{(s)}_{\ts+1} = \begin{bmatrix}
    \modelparaest^{(s)}\\
    f_{\modelparaest^{(s)}}(\widehat{\stats}^s_{\ts},\inp_{\ts})
\end{bmatrix}
\end{equation*}
and
\begin{equation}
    \augpara_\ts^{(s)} = \begin{cases}
        \augparaest_{\ts}&\text{for}\quad s=0\\
        \augparaest_{\ts} +  \eta^{1/2}[\errorcovjoint_{\ts}^{1/2}]_s &\text{for}\quad s=1,\dots,d_{\augpara}\\
        \augparaest_{\ts} -  \eta^{1/2}[\errorcovjoint_{\ts}^{1/2}]_s&\text{for}\quad s=d_{\augpara}+1,\dots,2d_{\augpara}
    \end{cases}.
\end{equation}
The weights $w^{(s)}$ and $\eta$ are given in \cite{julier2004unscented,menegaz2015systematization,candy2016bayesian}. The covariance matrix $\errorcovjoint_{\ts}$ is computed similarly via outer products of the samples. Eq. \eqref{eq:sigmapoint_transformation} is applied sequentially to produce running predictions of $\stats_{\ts}$ for use in \eqref{eq:optimize-relax}.

\subsection{Adaptive design method}

Using the joint estimate $(\modelparaest, \statsest_{\ts}, \noisevest)$ described above, we can finally evaluate the criterion in \eqref{eq:optimize-relax}. This design problem can be converted into an unconstrained form by transforming the inputs $\cinp \rightarrow \mathbb{R}^{d_u}$ as follows: \begin{equation}
    \widetilde{\kinp}_{\ts} = [\widetilde{u}(\inp_{\ts}),\dots,\widetilde{u}(\inp_{\ts+\hs-1})]
\label{eq:transformedinputs}
\end{equation}
where $\widetilde{u}(\inp_\ts) = g\big((\inp_\ts-\inp_\text{min})\oslash (\inp_\text{max}-\inp_\text{min})\big)$ 
and where $g$ is a differentiable and invertible function, e.g., the element-wise inverse sigmoid function  $[g(z)]_i=\ln{\frac{z_i}{1-z_i}} $. 

Thus \eqref{eq:transformedinputs} is unconstrained and computationally convenient for adaptive experiment design. The map $\widetilde{\kinp}_{\ts}$ recovers the original inputs by:
 \begin{equation}
     \kinp_{\ts} = [\widetilde{\inp}^{-1}(\widetilde{\inp}_{\ts}),\dots,\widetilde{\inp}^{-1}(\widetilde{\inp}_{\ts+\hs-1})]
    \label{eq:U_inversemap}
 \end{equation}
where $\widetilde{\inp}^{-1}(\Tilde{\inp}_\ts) =  g^{-1}(\Tilde{\inp}_\ts)\odot(\inp_{\max}-\inp_{\min}) + \inp_{\min}$ and $[g^{-1}(z)]_i = \frac{1}{1+e^{-z_i}}$. Using the transformed inputs $\widetilde{\kinp}_{\ts}$, by substituting \eqref{eq:U_inversemap} in \eqref{eq:optimize-relax}, results in an unconstrained optimization problem. At time $\ts$, the input 
$$\inp^*_{\ts} = \widetilde{u}^{-1}(\inp^*_{\ts})$$
is computed and applied in the subsequent time step.

The adaptive input design and online estimation method can now be summarized as follows:
\begin{enumerate}
    \item Observe $\blocksize$ new samples, solve \eqref{eq:blockwiseestimator} and compute \eqref{eq:estcov}.
    \item Update estimate of $\stats_{\ts}$ and $\errorcovjoint_{\ts}$ using \eqref{eq:sigmapoint_transformation}.
    \item Solve \eqref{eq:optimize-relax} in unconstrained form and apply input $\inp^*_{\ts}$.
    \item Update $\ts:=\ts+1$
    \item Go to 1) if $\ts$ is a multiple of $\blocksize$, otherwise go to 2).
\end{enumerate}

To solve the non-convex minimization problem in \eqref{eq:blockwiseestimator}, we employ a numerical optimization strategy suitable for rugged cost surfaces. Specifically, we utilize the Basin Hopping algorithm \cite{leary2000global,olson2012basin} to locate an approximate solution to \eqref{eq:blockwiseestimator}, then using a quasi-Newton limited-memory BFGS algorithm \cite{liu1989limited} for local optimization. To solve the design problem \eqref{eq:optimize-relax}, we leverage an available automatic differentiation method \cite{pytorch} alongside the Adam optimizer \cite{adam}, using a constant stepsize of $10^{-3}$. The code implementations are available online.\footnote{\url{https://github.com/jingwei91hu/adaptive-input-design}} The runtime of the design method using a standard laptop with single thread CPU is in the range of 0.2 to 5 seconds, depending on $\hs$ and model complexity.

%The online receding horizon problem \eqref{eq:optimize-relax} is tackled using a first-order gradient method TODO.

\section{Numerical Experiments}   
In the experiments presented below, we use a nonlinear pendulum model as an example of \eqref{eq:nonlinearsystem}:
\begin{equation}
\begin{split}
         \stats_{\ts+1}^{(1)} &= \stats_{\ts}^{(1)}+\stats_{\ts}^{(2)}\dt  \\
         \stats_{\ts+1}^{(2)} &= \stats_{\ts}^{(2)}+ (\modelpara_1\sin{\stats_{\ts}^{(1)}} + \modelpara_2\inp_{\ts})\dt,
         \label{eq:pendulum}
\end{split}
\end{equation}
where $\stats_{\ts}^{(1)}$ represents the pendulum's angular displacement and $\stats_{\ts}^{(2)}$ denotes its angular velocity $\inp_{\ts}$. Additional examples are provided in the Supplementary Materials.

The experiments are initialized with a block of $\blocksize$ samples generated from \prbstext{} inputs. The initial prior estimates, $\modelparaest_0$ and $\statsest_{1|0}$, are randomly drawn from a Gaussian distribution with zero mean and chose to be uninformative by assigning a large covariance matrix $\errorcovjoint_{1|0} = 10^4 \cdot I$. The initial noise estimates $\noisevest_0$ are presumed to be accurate, and their covariance matrix $\errorcovv_0$ is configured to correspond to a 0.1\% uncertainty in the noise standard deviations. For all the experiments, the weight in \eqref{eq:optimize-relax} was set to $\gamma = 400 \gg d_{\theta}$, where $d_{\theta}$ equals 2 or 3 below.

To provide a succinct evaluation of the parameter estimates, we use the normalized mean squared error 
\begin{equation*}
  \textsc{mse}_{\ts} = \sum^{d_{\modelpara}}_{k=1} \frac{\expectation[|\modelpara_k - \modelparaest_{k,\ts}|^2]}{\modelpara^2_k} 
\end{equation*}
and the corresponding (approximate) Cram\'{e}r-Rao bounds (\textsc{crb}) obtained by evaluating the \fimtext{} in \eqref{eq:fim} at the true parameters. In addition, we quantify the (relative) violation of operational constraints \eqref{eq:stateconstraints} using:
\begin{equation*}
\textsc{ocv}_t = \expectation\left[\|(S\stats_{\min}-S\stats_{\ts})^{+} + (S\stats_{\ts}- S\stats_{\text{max}})^{+})\|\right] ,
\end{equation*}
with $S$ from \eqref{costoperationalconstraints}.

\begin{figure}
   \centering
    \includegraphics[width=0.8\linewidth]{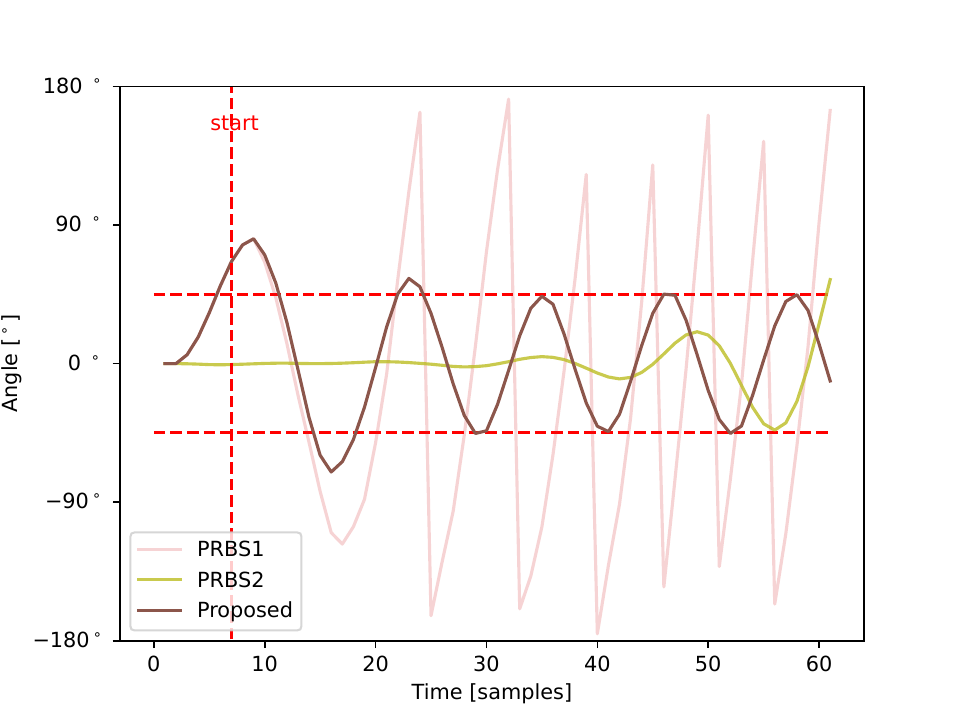}
    \vspace{-0.5cm}
    \caption{Example realization of pendulum system. Angle $x_\ts^{(1)}$ when applying two different \prbstext{} inputs and the adaptive design that steers the angle within the interval $[-45^\circ, 45^\circ]$.}
    \label{fig:invertedpendulum_output_c}
    \end{figure}

%\subsection{Pendulum system}
%\label{sec:pendulum}

\begin{figure}[!ht]
    \centering
    \includegraphics[width=0.8\linewidth]{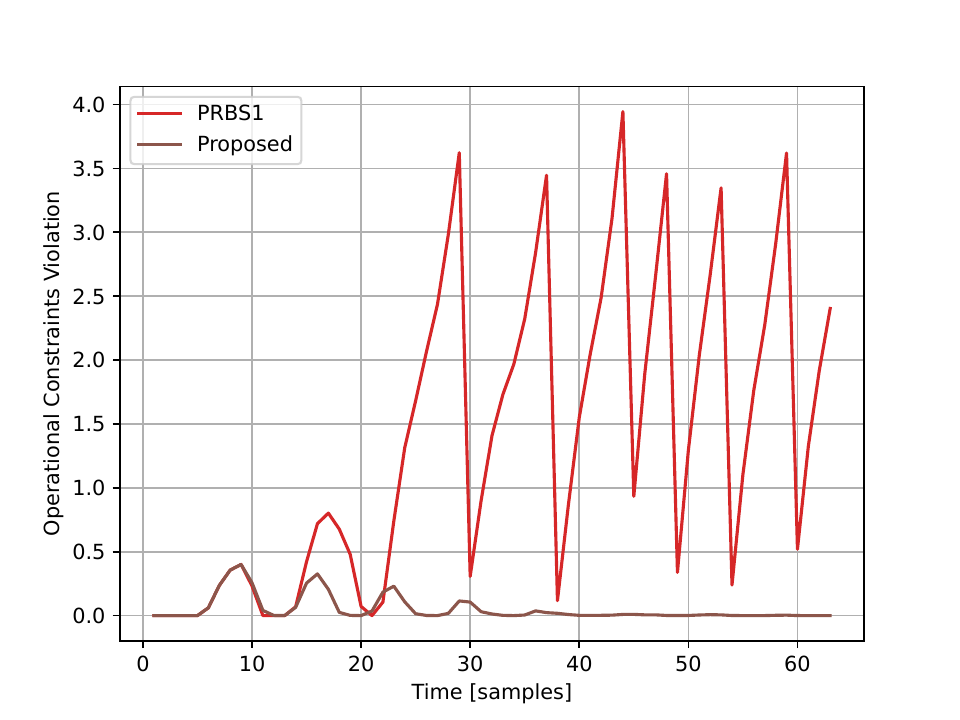}
    \vspace{-0.5cm}
    \caption{Relative violation of operational constraint \eqref{eq:stateconstraints}, evaluated using 100 Monte Carlo runs.}
    \label{fig:pendulum-violation}
\end{figure}

\begin{figure}[!ht]
    \centering
    \includegraphics[width=0.8\linewidth]{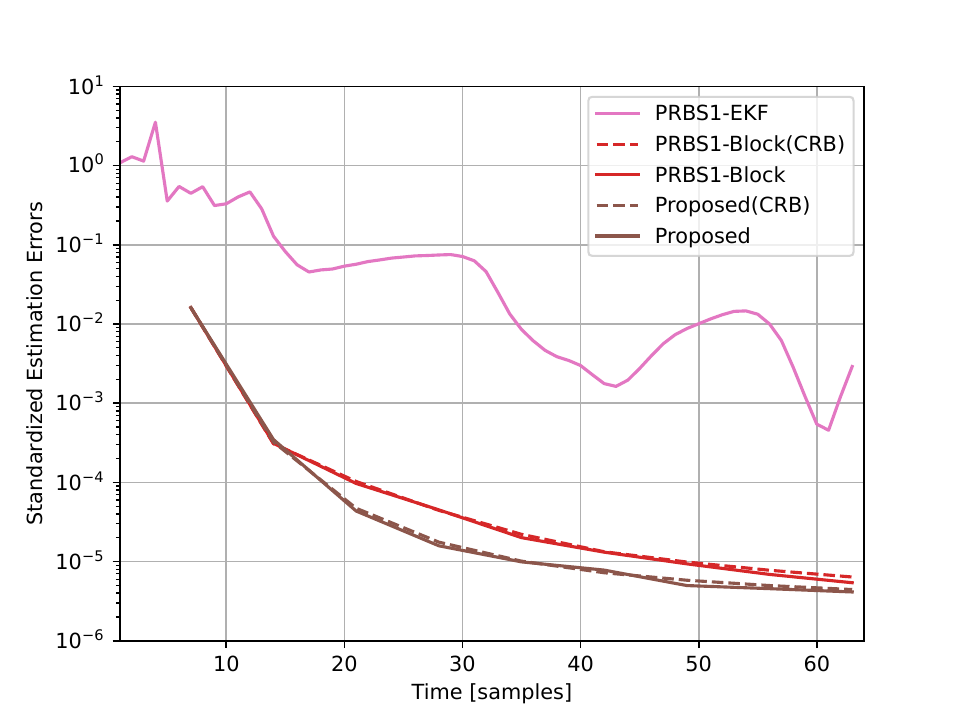}
    \vspace{-0.5cm}
    \caption{Mean-square errors of parameter estimator along with (approximate) Cram\'{e}r-Rao bounds. The estimation errors for \prbstext{}2 were too large to be included in the plot. Results are based on 100 Monte Carlo runs.}
    \label{fig:invertedpendulum_res}
\end{figure}

We set the  model parameters to $\modelpara_1=-24$, determined by the gravitational acceleration, and $\modelpara_2=1$, related to the pendulum's length. The sampling period is  $\Delta T = 0.1$ s. Only the angular displacement is observed, meaning $\regressormatrix = [1 \; 0]$ in \eqref{eq:linearobservation}, with standard deviation of noise $\sqrt{\noisev} = 0.01$ rad. The system is initialized at $\stats_0 = [0 \; 0]^\top$ with the goal of gathering informative samples while keeping the pendulum within  $[\stats^{(1)}_{\min},\stats^{(1)}_{\max}] = [- 45^\circ, + 45^\circ]$. The input constraint was $|\inp_{\ts}| \leq 10$. In the experiment, for the adaptive design, we use horizon length $\hs = 6$ and block size $\blocksize = 7$.

Figure~\ref{fig:invertedpendulum_output_c} provides an illustration of three experiments: The first is \prbstext{}1 which only satisfies input amplitude constraints, whereas the second one, \prbstext{}2, is tuned to the best of our ability to satisfy the operational constraints during an experiment. In contrast, the adaptive design steers the angular state within its constraints. A more systematic comparison of constraint violations is provided in Fig.~\ref{fig:pendulum-violation}. 

Finally, Figure~\ref{fig:invertedpendulum_res} shows the estimation errors and \textsc{crb}s. We see that the   online estimator outperforms the Extended Kalman Filter (\textsc{ekf}), with parameter-augmented states, when using the \prbstext{}1 design. We also note that the proposed design lowers the bound while steering within the state constraints. The resulting errors of the online errors are correspondingly lower.

\section{Conclusion}

We have proposed an adaptive experiment design method that steers the system within system constraints and an online estimator for nonlinear systems that provides parameter estimates and necessary state estimates. We demonstrate the ability of the method to perform safer experiments online, while steering the system within operational constraints, yet yield substantial gains in estimation accuracy and, therefore, reduced experiment time.

\newpage
\bibliographystyle{IEEEtranS}
\bibliography{ref}

\end{document}

%% file: math_variables.tex
\newcommand{\ts}{t}
\newcommand{\tp}{\tau}
\newcommand{\hs}{k}

\newcommand{\modelpara}{\theta}

\newcommand{\augpara}{z}

\newcommand{\loglikelihood}{\ell}
\newcommand{\loglikelihoodapprox}{\widetilde{\loglikelihood}}
\newcommand{\FIM}{J}
 \newcommand{\FIMapprox}{\widetilde{J}}
\newcommand{\expectation}{\mathbb{E}}

\newcommand{\cost}{\mathcal{J}}
\newcommand{\inp}{u}
\newcommand{\kinp}{U}
\newcommand{\cinp}{\mathcal{U}}

\newcommand{\oup}{y}

\newcommand{\stats}{x}

\newcommand{\cstats}{\mathcal{X}}
\newcommand{\prederr}{\varepsilon}

\newcommand{\blocksize}{b}

\newcommand{\noisecov}{V}
\newcommand{\noisev}{v}
\newcommand{\regressormatrix}{H}
\newcommand{\errorcov}{C}
\newcommand{\errorcovjoint}{P}
\newcommand{\errorcovv}{Q}
\newcommand{\errorcovapprox}{\widetilde{\errorcov}}
\newcommand{\augparaest}{\widehat{\augpara}}
\newcommand{\statsest}{\widehat{\stats}}

\newcommand{\modelparaest}{\widehat{\modelpara}}

\newcommand{\noisevest}{\widehat{\noisev}}
\newcommand{\prederrgrad}{E}

\newcommand{\dt}{\Delta{T}}

\DeclareMathOperator*{\argmin}{arg\,min}

\newcommand{\fimtext}{\textsc{fim}}
\newcommand{\prbstext}{\textsc{prbs}}
\newcommand{\mletext}{\textsc{mle}}